\documentclass[aps,prl,reprint,groupedaddress]{revtex4-1}
\usepackage{graphicx} 
\usepackage{dcolumn} 
\usepackage{bm}
\usepackage{hyperref}

\begin{document}

\title{Dynamics of pion production in heavy-ion collisions around 1A GeV energies}

\author{Zhao-Qing Feng}
\email{fengzhq@impcas.ac.cn}
\author{Gen-Ming Jin}
\affiliation{Institute of Modern Physics, Chinese Academy of
Sciences, Lanzhou 730000, People's Republic of China}

\date{\today}

\begin{abstract}
Within the framework of the improved isospin dependent quantum
molecular dynamics (ImIQMD) model, the dynamics of pion emission in
heavy-ion collisions in the region of 1 A GeV energies as a probe of nuclear
symmetry energy at supra-saturation densities is investigated
systematically. The total pion multiplicities and the
$\pi^{-}/\pi^{+}$ yields are calculated for selected Skyrme
parameters SkP, SLy6, Ska and SIII, and also for the cases of
different stiffness of symmetry energy with the parameter SLy6.
The influence of Coulomb potential, symmetry energy and in-medium pion potential
on the pion production is investigated and compared each other by analyzing the distributions
of transverse momentum and longitudinal rapidity and also the excitation functions
of the total pion and the $\pi^{-}/\pi^{+}$ ratio. The directed flow, elliptic flow
and polar angle distributions are calculated at the cases of different collision centralities
and also the various stiffness of the symmetry energies. A comparison of the calculations with the available experimental data is performed.
\begin{description}
\item[PACS number(s)]
25.75.-q, 13.75.Gx, 25.80.Ls
\end{description}
\end{abstract}

\maketitle

\section{I. Introduction}

Heavy-ion collisions induced by radioactive beam at intermediate
energies play a significant role to extract the information of
nuclear equation of state (EoS) of isospin asymmetric nuclear matter
under extreme conditions, such as high density, high temperature and large isospin asymmetry etc.
Besides nucleonic observables such as
rapidity distribution and flow of free nucleons or light clusters
(such as deuteron, triton and alpha etc.), also mesons emitted from
the reaction zone can be probes of the hot and dense nuclear matter.
The energy per nucleon in the isospin asymmetric nuclear matter is
usually expressed as
$E(\rho,\delta)=E(\rho,\delta=0)+E_{\textrm{sym}}(\rho)\delta^{2}+\textsc{O}(\delta^{2})$
in terms of baryon density $\rho=\rho_{n}+\rho_{p}$, relative
neutron excess $\delta=(\rho_{n}-\rho_{p})/(\rho_{n}+\rho_{p})$,
energy per nucleon in a symmetric nuclear matter $E(\rho,\delta=0)$
and bulk nuclear symmetry energy
$E_{\textrm{sym}}=\frac{1}{2}\frac{\partial^{2}E(\rho,\delta)}{\partial
\delta^{2}}\mid_{\delta=0}$. In general, two different forms have
been predicted by some microscopical or phenomenological many-body
approaches. One is the symmetry energy increases monotonically with the baryon
density, and the other is the symmetry energy increases initially up
to a supra-saturation density and then decreases at higher
densities. The difference of the symmetry energy at supra-saturation densities predicted by transport models is huge.
It is not only in understanding the reaction dynamics, the high-density behavior of the symmetry energy also has an important application in astrophysics,
such as the structure of neutron star, the cooling of protoneutron stars, the nucleosynthesis
during supernova explosion of massive stars etc \cite{St05}.
With the establishment of high-energy radioactive beam
facilities in the world, such as the CSR (IMP in Lanzhou, China),
FAIR (GSI in Darmstadt, Germany), RIKEN (Japan), SPIRAL2 (GANIL in
Caen, France) and FRIB (MSU, USA) \cite{Li08}, the high-density
behavior of the symmetry energy can be studied more detail
experimentally in the near future.

Based on recent analysis of experimental data associated
with transport models, a symmetry energy of the form
$E_{\textrm{sym}}(\rho)\approx 31.6(\rho/\rho_{0})^{\gamma}$ MeV
with $\gamma=0.69-1.05$ was extracted for densities between
0.1$\rho_{0}$ and 1.2$\rho_{0}$ \cite{Li08,Ch05}. Theoretically, the symmetry
energy at supra-saturation densities can be investigated by
analyzing isospin sensitive observables that are emitted in the high-density region, such as
the neutron/proton ratio of emitted nucleons at mid-rapidity, $\pi^{-}/\pi^{+}$,
$\Sigma^{-}/\Sigma^{+}$ and $K^{0}/K^{+}$ ratios etc \cite{Li08}. Heavy-ion collisions around 1A GeV
energies can reach a 2-3$\rho_{0}$ nuclear matter. The emission of pion
in heavy-ion collisions in the region 1 A GeV is especially sensitive as a probe
of symmetry energy at supra-saturation densities. Recently, a
very soft symmetry energy at supra-saturation densities was pointed out
through fitting the FOPI data \cite{Re07} by using IBUU04 model \cite{Xi09}. The inverse results
were also reported by the Catania group with the RBUU model, in which a hard symmetry energy in
the high-density region resulted in the large values of the $\pi^{-}/\pi^{+}$ and $K^{0}/K^{+}$ ratios \cite{Fe06,Pr10}.
Further investigations of the pion emissions in the 1 A GeV region are still
necessary by improving transport models or developing some new
approaches. The ImIQMD model has been successfully applied to treat
dynamics in heavy-ion fusion reactions near Coulomb barrier and also to describe the capture of two heavy colliding nuclides
\cite{Fe05,Fe08}. Some further improvements of the ImIQMD model have been performed in order to
investigate the pion dynamics in heavy-ion collisions \cite{Fe09,Fe10}.

A systematic comparison of the ImIQMD results and the available experimental data is performed in this work. Calculations are focused on the pion production in heavy-ion collisions around 1A GeV energies. Some observables sensitive to the high-density behavior of the symmetry energy are investigated and discussed. The paper is organized as follows. In Sec. II we give a detailed description of the ImIQMD model. Calculated
results of pion dynamics in heavy-ion collisions and constraining the symmetry energy at supra-saturation densities
are given in Sec. III. In Sec. IV conclusions are discussed.

\section{II. Model description}

The same as the QMD \cite{Ai91} or IQMD model \cite{Ha98,Ch98}, the wave function for each
nucleon in ImIQMD is represented by a Gaussian wave packet
\begin{eqnarray}
\psi_{i}(\mathbf{r},t)=&& \frac{1}{(2\pi
L)^{3/4}}\exp\left[-\frac{(\mathbf{r}-\mathbf{r}_{i}(t))^{2}}{4L}\right]  \nonumber \\
&& \times \exp\left(\frac{i\mathbf{p}_{i}(t)\cdot\mathbf{r}}{\hbar}\right).
\end{eqnarray}
Here $\mathbf{r}_{i}(t)$, $\mathbf{p}_{i}(t)$ are the centers of the
$i$th nucleon in the coordinate and momentum space, respectively.
The $L$ is the square of the Gaussian wave packet width, which
depends on the mass number of nucleus. The total N-body wave function is
assumed as the direct product of the coherent states, where the
anti-symmetrization is neglected. After performing Wigner transformation for Eq.
(1), we get the Wigner density as
\begin{equation}
f(\mathbf{r},\mathbf{p},t)=\sum_{i}f_{i}(\mathbf{r},\mathbf{p},t)
\end{equation}
with
\begin{eqnarray}
f_{i}(\mathbf{r},\mathbf{p},t)=&& \frac{1}{(\pi\hbar)^{3}}\exp  \nonumber \\
&& \times \left[-
\frac{(\mathbf{r}-\mathbf{r}_{i}(t))^{2}}{2L}-\frac{(\mathbf{p}-\mathbf{p}_{i}(t))^{2}\cdot
2L}{\hbar^{2}}\right].
\end{eqnarray}
The density distributions in coordinate and momentum space are given
by
\begin{eqnarray}
\rho(\mathbf{r},t)=&& \int f(\mathbf{r},\mathbf{p},t) d\mathbf{p}  \nonumber \\
=&& \sum_{i}\frac{1}{(2\pi L)^{3/2}}\exp\left[-
\frac{(\mathbf{r}-\mathbf{r}_{i}(t))^{2}}{2L}\right],
\end{eqnarray}
\begin{eqnarray}
g(\mathbf{p},t)=&& \int f(\mathbf{r},\mathbf{p},t) d\mathbf{r}  \nonumber \\
=&& \sum_{i}\left(\frac{2L}{\pi\hbar^{2}}\right)^{3/2}\exp\left[-
\frac{(\mathbf{p}-\mathbf{p}_{i}(t))^{2}\cdot 2L}{\hbar^{2}}\right],
\end{eqnarray}
respectively, where the sum runs over all nucleons in the reaction systems.

The time evolutions of the baryons and pions in
the system under the self-consistently generated mean-field are
governed by Hamilton's equations of motion, which read as
\begin{eqnarray}
\dot{\mathbf{p}}_{i}=-\frac{\partial H}{\partial\mathbf{r}_{i}},
\quad \dot{\mathbf{r}}_{i}=\frac{\partial
H}{\partial\mathbf{p}_{i}}.
\end{eqnarray}
Here we omit the shell correction part in the Hamiltonian $H$ as
described in Ref. \cite{Fe08}. The Hamiltonian of baryons consists
of the relativistic energy, the effective interaction potential energy and
the momentum dependent part as follows:
\begin{equation}
H_{B}=\sum_{i}\sqrt{\textbf{p}_{i}^{2}+m_{i}^{2}}+U_{int}+U_{mom}.
\end{equation}
Here the $\textbf{p}_{i}$ and $m_{i}$ represent the momentum and the
mass of the baryons.

The effective interaction potential is composed of the Coulomb
interaction and the local interaction
\begin{equation}
U_{int}=U_{Coul}+U_{loc}.
\end{equation}
The Coulomb interaction potential is calculated by
\begin{equation}
U_{Coul}=\frac{1}{2}\sum_{i,j,j\neq
i}\frac{e_{i}e_{j}}{r_{ij}}erf(r_{ij}/\sqrt{4L})
\end{equation}
where the $e_{j}$ is the charged number including protons and
charged resonances. The $r_{ij}=|\mathbf{r}_{i}-\mathbf{r}_{j}|$ is
the relative distance of two charged particles.

The local interaction potential energy is derived directly from the Skyrme
energy-density functional and expressed as
\begin{equation}
U_{loc}=\int V_{loc}(\rho(\mathbf{r}))d\mathbf{r}.
\end{equation}
The local potential energy-density functional reads
\begin{eqnarray}
V_{loc}(\rho)=&& \frac{\alpha}{2}\frac{\rho^{2}}{\rho_{0}}+
\frac{\beta}{1+\gamma}\frac{\rho^{1+\gamma}}{\rho_{0}^{\gamma}}+
\frac{g_{sur}}{2\rho_{0}}(\nabla\rho)^{2}  \nonumber \\
&& + \frac{g_{sur}^{iso}}{2\rho_{0}}[\nabla(\rho_{n}-\rho_{p})]^{2}  \nonumber \\
&& + \left(a_{sym}\frac{\rho^{2}}{\rho_{0}}+b_{sym}\frac{\rho^{1+\gamma}}{\rho_{0}^{\gamma}}+
c_{sym}\frac{\rho^{8/3}}{\rho_{0}^{5/3}}\right)\delta^{2}  \nonumber \\
&& + g_{\tau}\rho^{8/3}/\rho_{0}^{5/3},
\end{eqnarray}
where the $\rho_{n}$, $\rho_{p}$ and $\rho=\rho_{n}+\rho_{p}$ are
the neutron, proton and total densities, respectively, and the
$\delta=(\rho_{n}-\rho_{p})/(\rho_{n}+\rho_{p})$ is the isospin
asymmetry. Here, all the terms in the Skyrme energy functional are included in the model besides the spin-orbit coupling. The coefficients $\alpha$, $\beta$, $\gamma$, $g_{sur}$,
$g_{sur}^{iso}$, $g_{\tau}$ are related to the Skyrme parameters
$t_{0}, t_{1}, t_{2}, t_{3}$ and $x_{0}, x_{1}, x_{2}, x_{3}$ as \cite{Fe08},
\begin{eqnarray}
&& \frac{\alpha}{2}=\frac{3}{8}t_{0}\rho_{0}, \quad
\frac{\beta}{1+\gamma}=\frac{t_{3}}{16}\rho_{0}^{\gamma}, \\
&& \frac{g_{sur}}{2}=\frac{1}{64}(9t_{1}-5t_{2}-4x_{2}t_{2})\rho_{0}, \\
&& \frac{g_{sur}^{iso}}{2}=-\frac{1}{64}[3t_{1}(2x_{1}+1)+t_{2}(2x_{2}+1)]\rho_{0}, \\
&& g_{\tau}=\frac{3}{80}\left(\frac{3}{2}\pi^{2}\right)^{2/3}(3t_{1}+5t_{2}+4x_{2}t_{2})\rho_{0}^{5/3}.
\end{eqnarray}
The parameters of the potential part in the bulk symmetry
energy term are also derived directly from Skyrme energy-density
parameters as
\begin{eqnarray}
&& a_{sym}=-\frac{1}{8}(2x_{0}+1)t_{0}\rho_{0}, \quad b_{sym}=-\frac{1}{48}(2x_{3}+1)t_{3}\rho_{0}^{\gamma}, \nonumber \\
&& c_{sym}=-\frac{1}{24}\left(\frac{3}{2}\pi^{2}\right)^{2/3}\rho_{0}^{5/3}[3t_{1}x_{1}-t_{2}(5x_{2}+4)].
\end{eqnarray}

The momentum dependent term in the Hamiltonian is taken as the same form in Ref. \cite{Ai87} and expressed as
\begin{equation}
U_{mom}=\frac{\delta}{2}\sum_{i,j,j\neq
i}\frac{\rho_{ij}}{\rho_{0}}[\ln(\epsilon(\textbf{p}_{i}-\textbf{p}_{j})^{2}+1)]^{2},
\end{equation}
with
\begin{equation}
\rho_{ij}=\frac{1}{(4\pi L)^{3/2}}\exp\left[
-\frac{(\textbf{r}_{i}-\textbf{r}_{j})^{2}}{4L}\right],
\end{equation}
which does not distinguish between protons and neutrons. The parameters $\delta$ and
$\epsilon$ were determined by fitting the real part of the
proton-nucleus optical potential as a function of incident energy from the experimental elastic scattering data.
The last two terms in Eq. (11) originate from the momentum-dependent part of the Skyrme interaction which, in principle, should contribute to the nucleon effective mass \cite{Br85,Ch97}. However, in the ImIQMD model their contribution is taken into account in the potential energy-density functional only and not in the momentum-dependent interaction given by the term of Eq. (17) solely. Therefore, the effective (Landau) mass in the model is fixed as $m_{\infty}^{\ast}=\left(\frac{1}{m}+\frac{1}{|\textbf{p}|}|\frac{dU_{mom}}{d\textbf{p}}|\right)^{-1}$ with the free mass $m$ at Fermi momentum $\textbf{p}=\textbf{p}_{F}$ \cite{Da05}, which produces the value of $m_{\infty}^{\ast}/m=0.77$ for all sets of the model parameters.

\begin{table*}
\caption{\label{tab:table3}ImIQMD parameters and properties of symmetric nuclear
matter for Skyrme effective interactions after the inclusion of the
momentum dependent interaction with parameters $\delta$=1.57 MeV and
$\epsilon$=500 c$^{2}$/GeV$^{2}$.}
\begin{ruledtabular}
\begin{tabular}{ccccccccc}
&Parameters                 &SkM*   &Ska    &SIII   &SVI    &SkP    &RATP   &SLy6   \\
\hline
&$\alpha$ (MeV)             &-325.1 &-179.3 &-128.1 &-123.0 &-357.7 &-250.3 &-296.7 \\
&$\beta$  (MeV)             &238.3  &71.9   &42.2   &51.6   &286.3  &149.6  &199.3  \\
&$\gamma$                   &1.14   &1.35   &2.14   &2.14   &1.15   &1.19   &1.14   \\
&$g_{sur}$(MeV fm$^{2}$)    &21.8   &26.5   &18.3   &14.1   &19.5   &25.6   &22.9   \\
&$g_{sur}^{iso}$(MeV fm$^{2}$)&-5.5 &-7.9   &-4.9   &-3.0   &-11.3  &0.0    &-2.7   \\
&$g_{\tau}$ (MeV)           &5.9    &13.9   &6.4    &1.1    &0.0    &11.0   &9.9    \\
&$C_{sym}$ (MeV)            &30.1   &33.0   &28.2   &27.0   &30.9   &29.3   &32.0   \\
&$a_{sym}$ (MeV)            &62.4   &29.8   &38.9   &42.9   &94.0   &79.3   &130.6  \\
&$b_{sym}$ (MeV)            &-38.3  &-5.9   &-18.4  &-22.0  &-63.5  &-58.2  &-123.7 \\
&$c_{sym}$ (MeV)            &-6.4   &-3.0   &-3.8   &-5.5   &-13.0  &-4.1   &12.8   \\
&$\rho_{\infty}$ (fm$^{-3}$)&0.16   &0.155  &0.145  &0.144  &0.162  &0.16   &0.16   \\
&$K_{\infty}$ (MeV)         &215    &262    &353    &366    &200    &239    &230    \\
\end{tabular}
\end{ruledtabular}
\end{table*}

In Table 1 we list the ImIQMD parameters related to several typical
Skyrme forces after including the momentum dependent interaction.
The parameters $\alpha$, $\beta$ and $\gamma$ are redetermined in
order to reproduce the binding energy ($E_{B}$=-16 MeV) of symmetric
nuclear matter at saturation density $\rho_{0}$ and to satisfy the saturation properties by
relation $\frac{\partial E/A}{\partial\rho}\mid _{\rho=\rho_{0}}$=0
for a given incompressibility. Combined Eq. (16) with the kinetic
energy part, the symmetry energy per nucleon in the ImIQMD model is
given by
\begin{eqnarray}
E_{sym}(\rho)=&& \frac{1}{3}\frac{\hbar^{2}}{2m}\left(\frac{3}{2}\pi^{2}\rho\right)^{2/3}+
a_{sym}\frac{\rho}{\rho_{0}}+b_{sym}\left(\frac{\rho}{\rho_{0}}\right)^{\gamma}   \nonumber \\
&& + c_{sym}\left(\frac{\rho}{\rho_{0}}\right)^{5/3}.
\end{eqnarray}
More clearly compared with other transport models, such as IBUU04, RBUU, and IQMD etc, the bulk symmetry
energy can be expressed as
\begin{equation}
E_{sym}(\rho)=\frac{1}{3}\frac{\hbar^{2}}{2m}\left(\frac{3}{2}\pi^{2}\rho\right)^{2/3}+
\frac{1}{2}C_{pot}\left(\frac{\rho}{\rho_{0}}\right)^{\gamma_{s}}.
\end{equation}
The value $\gamma_{s}=1$ is used in the IQMD model \cite{Ha98,Ch98}. In
Fig. 1 we show a comparison of the nuclear symmetry energy at the situations of
different Skyrme forces SkP, SLy6, Ska and SIII calculated by
Eq. (19), and also the cases in Eq. (20) of $\gamma_{s}$= 0.5 (soft), 1 (linear), 2 (hard), 3 (superhard) with the coefficient $C_{pot}$=38 MeV, which have the value of the symmetry energy $E_{sym}(\rho_{0})$=31.5 MeV at the saturation nuclear density $\rho_{0}$=0.165 fm$^{-3}$. The parameter SLy6 is used in the calculations by default.

\begin{figure}
\includegraphics[width=8 cm]{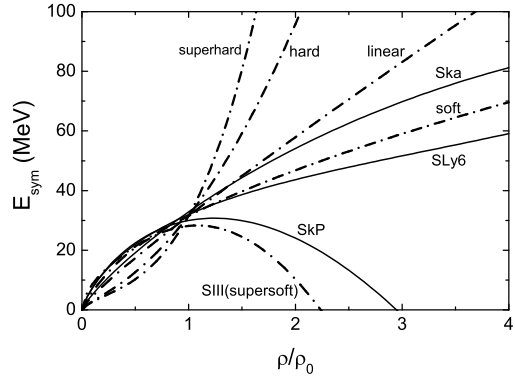}
\caption{\label{fig:epsart} The density dependence of the nuclear
symmetry energy for different Skyrme forces SkP, Sly6, Ska and SIII (supersoft),
and for the cases of soft, linear, hard and superhard trends.}
\end{figure}

The pion is created by the decay of the resonances $\triangle$(1232)
and N*(1440) which are produced in inelastic NN scattering. The
reaction channels are given as follows:
\begin{eqnarray}
&& NN \leftrightarrow N\triangle, \quad  NN \leftrightarrow NN^{\ast}, \quad  NN
\leftrightarrow \triangle\triangle,  \nonumber \\
&& \Delta \leftrightarrow N\pi, \quad  N^{\ast} \leftrightarrow N\pi, \quad NN \rightarrow NN\pi (\textrm{s-state}).
\end{eqnarray}
The cross sections of each channel to produce resonances are
parameterized by fitting the data calculated with the one-boson
exchange model \cite{Hu94}. In the 1 A GeV region, there are mostly
$\Delta$ resonances which disintegrate into a $\pi$ and a nucleon,
however, the $N^{\ast}$ yet gives considerable contribution to the
high energetic pion yield. The energy and momentum dependent decay
width is used in the calculation \cite{Fe09}.

Analogously to baryons, the evolution of pions is also determined by the Hamiltonian, which is given by
\begin{eqnarray}
H_{\pi}&& = \sum_{i=1}^{N_{\pi}}\left( V_{i}^{\textrm{Coul}} + \omega(\textbf{p}_{i},\rho_{i}) \right)   \nonumber \\
&& = \sum_{i=1}^{N_{\pi}}\left( V_{i}^{\textrm{Coul}} + \sqrt{\textbf{p}_{i}^{2}+m_{\pi}^{2}}
+ \textrm{Re}V_{\pi}^{opt}(\textbf{p}_{i},\rho_{i}) \right),
\end{eqnarray}
where the $\textbf{p}_{i}$ and $m_{\pi}$ represent the momentum and
the mass of the pions. The Coulomb interaction is given by
\begin{equation}
V_{i}^{\textrm{Coul}}=\sum_{j=1}^{N_{B}}\frac{e_{i}e_{j}}{r_{ij}},
\end{equation}
where the $N_{\pi}$ and $N_{B}$ are the total numbers of pions and
baryons including charged resonances. The pion optical potential $\textrm{Re}V_{\pi}^{opt}$ originates
from the medium effects in the hot and dense nuclear matter. In the calculation, we can also choose that of the vacuum, i.e.,
the $\textrm{Re}V_{\pi}^{opt}$ is set equal zero. The influence of the pionic mean field in heavy-ion collisions on the transverse momentum distribution
was investigated by using a phenomenological ansatz and a microscopic approach based on the $\Delta$-hole model by Fuchs \emph{et al.} \cite{Fu97}.
Here we use the phenomenological ansatz suggested by Gale and Kapusta \cite{Ga87}. Then the dispersion relation reads
\begin{eqnarray}
\omega(\textbf{p}_{i},\rho_{i})=\sqrt{(|\textbf{p}_{i}|-p_{0})^{2}+m_{0}^{2}}-U,  \\
U=\sqrt{p_{0}^{2}+m_{0}^{2}}-m_{\pi},  \\
m_{0}=m_{\pi}+6.5(1-x^{10})m_{\pi},   \\
p_{0}^{2}=(1-x)^{2}m_{\pi}^{2}+2m_{0}m_{\pi}(1-x).
\end{eqnarray}
The phenomenological medium dependence on the baryon density is introduced via the coefficient $x(\rho_{i})=\exp(-a(\rho_{i}/\rho_{0}))$ with the parameter
$a=0.154$ and the saturation density $\rho_{0}$ in nuclear matter. Influence of the in-medium effects on the charged
pion ratio is also investigated in Ref. \cite{Xu10}.

\section{III. Results and discussions}

\subsection{A. Charge distributions}

Nuclear multifragmentation is a common phenomenon observed in
heavy-ion collisions at intermediate energies, in which one can
learn more about the properties of the fermionic nuclear matter,
such as the compressibility, the liquid-gas phase transition and
also the isospin asymmetric EoS. As a test of the ImIQMD model, we
calculated the charge distributions in multifragmentation reactions
of the system $^{197}$Au+$^{197}$Au at different incident energies
labeled in Fig. 2. The fragment multiplicity decreases with the
atomic number and the trend is more rapidly at the higher incident
energy The available experimental data \cite{De98,Re97} can
reproduced rather well besides the energy at 35 A MeV in the region of the intermediate mass fragments (IMFs) (Z=3-15).
The failure description of the production of the IMFs at several tens A MeV is due to the roughly treatment of fermionic nature in the dynamical evolution of nucleons.
These reactions are well described by the Statistical Multifragmentation Model (SMM) in which one assumes the formation of a thermal source and uses the statistical approach for its breakup \cite{Ag96}. Combination of transport models and SMM can reproduce the experimental data at fermionic energies. We constructed the fragments with the coalescence model, in which nucleons of the reaction system are
considered to belong to a cluster in the phase space with the
relative momentum smaller than $P_{0}$ and with the relative
distance smaller than $R_{0}$ (here $P_{0}$=200 MeV/c and
$R_{0}$=2.4 fm).

\begin{figure*}
\includegraphics{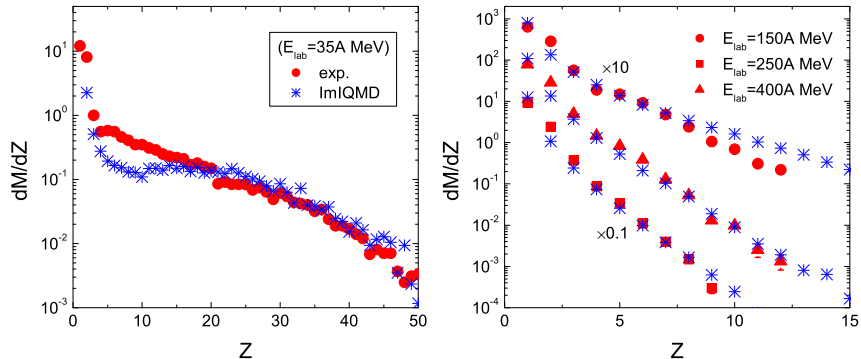}
\caption{\label{fig:wide} (Color online) Comparison of the
distribution of charged fragments in central $^{197}$Au+$^{197}$Au
collisions with the available experimental data at different
incident energies.}
\end{figure*}

\subsection{B. Production of the total pion and the $\pi^{-}/\pi^{+}$ yields}

Production of pion in heavy-ion collisions in the region of 1A GeV
is a primary product in nucleon-nucleon inelastic scattering and
mainly produced by the decay of the resonance $\triangle$(1232). So
it has enough yields to measure pion in experiments to get the
information of the EoS. Dynamics of the pion emission calculated by
transport models is helpful for understanding the experimental
observables. We calculated the time evolution of the multiplicities
of the total $\pi$, $\Delta(1232)$, $N^{\ast}(1440)$ and the ratio
$\rho/\rho_{0}$ in central $^{197}$Au+$^{197}$Au collisions at 1A
GeV incident energy as shown in Fig. 3. One can see that pions are
mainly in the domain at supra-saturation densities of compressed
nuclear matter larger than the normal density $\rho_{0}$. The
production of pions is also influenced by the Fermi motion of
baryons in the vicinity of the threshold energies. Figure 4 is a
comparison of the measured total pion multiplicity by the FOPI
collaboration in the reaction $^{40}$Ca+$^{40}$Ca for head on
collisions \cite{Re07} and the results calculated by the ImIQMD
model for Skyrme parameters SkP, SLy6, Ska and SIII in the left
panel, which correspond to different modulus of the
incompressibility as listed in table 1, and the ratio of the
calculated results to the experimental data at different incident
energies. The total multiplicity of pion is mainly determined by the
cross sections of the channels $NN \leftrightarrow N\triangle$. The
ImIQMD model with four Skyrme parameters predicts rather well the
total yields at higher incident energies, but slightly overestimates
the values near threshold energies, which may be influenced by the
in-medium cross sections. In the calculation we use the in-vacuum
cross sections of nucleon-nucleon elastic and inelastic collisions.
Reasonable consideration of the in-medium inelastic collisions in
producing $\Delta$ and $N^{\ast}$ is still an open problem in
transport models, which have been performed in Giessen-BUU model
\cite{La01}.

\begin{figure}
\includegraphics[width=8 cm]{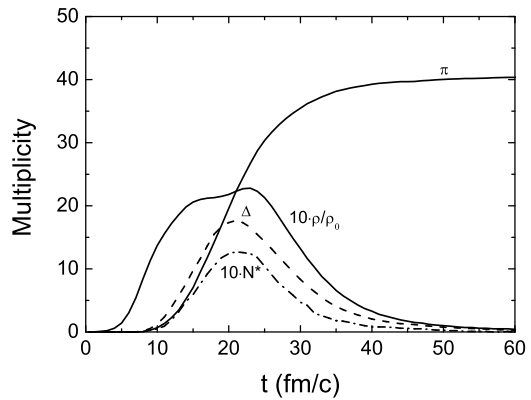}
\caption{\label{fig:epsart} Time evolution of the total $\pi$, $\Delta(1232)$, $N^{\ast}(1440)$ and the ratio $\rho/\rho_{0}$
in the reaction $^{197}$Au+$^{197}$Au for central collisions at incident energy 1A GeV.}
\end{figure}

\begin{figure*}
\includegraphics{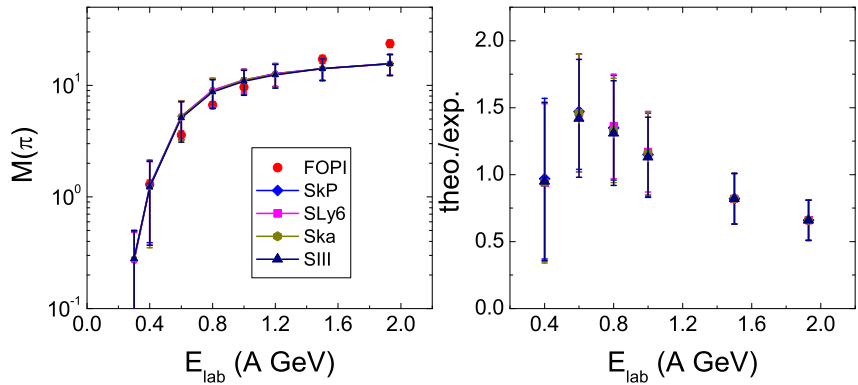}
\caption{\label{fig:wide} (Color online) The excitation functions of
the total pion multiplicities and the ratio of the calculated
results with different Skyrme parameters to the experimental data in
the reaction $^{40}$Ca+$^{40}$Ca for head on collisions.}
\end{figure*}

The $\pi^{-}$/$\pi^{+}$ ratio a sensitive probe to extract the
high-density behavior of the symmetry energy per energy. Using the
isobar model \cite{St86}, one gets the ratio $\pi^{-}$/$\pi^{+}$=2.09 for pions
from the $\Delta$ resonance, and $\pi^{-}$/$\pi^{+}$=1.8 from the N*
for the system $^{132}$Sn+$^{124}$Sn. These relations
are determined by the Clebsch-Gordan coefficients and globally valid, i.e. independent of the pion energy. On the
other hand, the statistical model predicts that the
$\pi^{-}$/$\pi^{+}$ ratio is sensitive to the difference in the
chemical potentials of neutrons and protons by the relation
$\pi^{-}/\pi^{+}\propto \exp [2(\mu_{n}-\mu_{p})/T]=\exp[8\delta
E_{sym}(\rho)/T]$, where the $T$ is the nuclear temperature \cite{Be80}.
The observed energy dependence of the $\pi^{-}$/$\pi^{+}$ ratio is
due to the re-scattering and absorption process of pions and
nucleons in the mean field of the compressed nuclear matter. We use
the free absorption cross sections in collisions of pions and
nucleons by fitting the experimental data. The branch ratio of the
charged $\pi$ and $\pi^{0}$ is determined by the Clebsch-Gordan
coefficients with the decay of the resonances $\triangle$(1232) and
N*(1440). The $\pi^{-}/\pi^{+}$ ratio is sensitive to the stiffness
of the symmetry energy at the lower incident energies. The
compressed nuclear matter with central density about two times of
the normal density is formed in heavy-ion collisions in the 1 A GeV
region. To extract more information of the symmetry energy in
heavy-ion collisions from the pion production, in Fig. 5 we
calculated the $\pi^{-}/\pi^{+}$ ratios as a function of the N/Z of
the Sn isotopes in the reactions $^{100}$Sn+$^{112}$Sn,
$^{112}$Sn+$^{112}$Sn, $^{112}$Sn+$^{124}$Sn, $^{124}$Sn+$^{124}$Sn
and $^{132}$Sn+$^{124}$Sn in the left panel, and the excitation
functions of the $\pi^{-}$/$\pi^{+}$ ratios with the force SLy6 for the coefficients by Eqs. (12)-(15) besides the symmetry energy term. Here, the symmetry energy is treated as the cases without the term or with but different stiffness of the potential part which corresponds to superhard ($\gamma_{s}$=3), hard
($\gamma_{s}$=2), linear ($\gamma_{s}$=1), soft ($\gamma_{s}$=0.5)
and supersoft (SIII). The symmetry energy reduces the
$\pi^{-}/\pi^{+}$ ratio for neutron-rich systems at the lower
incident energies because of its attractive force on protons and
opposite for neutrons in the dynamical evolution. So the probabilities of neutron-neutron collisions decrease after including the symmetry energy, which lead to the reduction of $\pi^{-}$ production. The ratio increases with the stiffness of the symmetry energy, but decreases for the case of the superhard symmetry energy. The phenomena can be explained from the fact that although the symmetry energy reduces the
N/Z ratio in the high-density region. A hard symmetry energy repulses neutrons of the colliding partners to flow the high-density region and
enhances the domain of the supra-saturation density. Therefore, the neutron-neutron collisions in the case of the hard symmetry energy take place in a larger region at supra-saturation densities than a soft one. A too hard symmetry energy also reduces the value of $\pi^{-}$/$\pi^{+}$ ratio because of the lower probabilities of the neutron-neutron collisions. The decrease of the $\pi^{-}$/$\pi^{+}$ ratio with the incident energy is mainly
owing to the production of pions from secondary nucleon-nucleon
collisions, such as a neutron converts a proton by producing
$\pi^{-}$. Subsequent collisions of the energetic proton can convert
again to neutron by producing $\pi^{+}$. The $\pi^{-}$/$\pi^{+}$ ratios of different situation of the symmetry energy at high incident energies (E$_{lab}>$1.0A GeV) are close each other and approach the value calculated by using the isobar model. The larger
$\pi^{-}$/$\pi^{+}$ ratio in the case of the hard symmetry energy was also
reported in Refs \cite{Fe06,Pr10}. Shown in Fig. 6 is the profile of pion production without the symmetry energy and with increasing the stiffness of the potential part of the symmetry energy in Eq. (20). The symmetry energy reduces the $\pi^{-}$ production owing to the decrease of the N/Z ratio in the high-density region. Influence of the stiffness of the symmetry energy on $\pi^{+}$ is negligible.

\begin{figure*}
\includegraphics{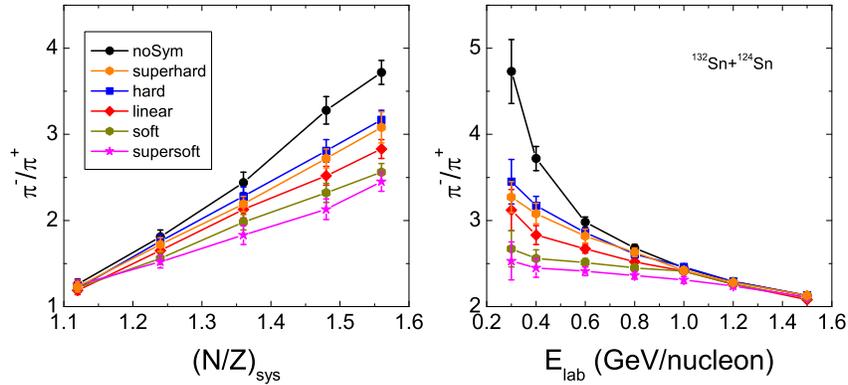}
\caption{\label{fig:wide} (Color online) The $\pi^{-}$/$\pi^{+}$
yields as functions of the neutron over proton N/Z of reaction
systems in collisions of Sn isotopes at energy E$_{lab}$=0.4A GeV and of the incident energy for the
reaction $^{132}$Sn+$^{124}$Sn for the cases of different stiffness of the symmetry energy and without symmetry energy.}
\end{figure*}

\begin{figure}
\includegraphics[width=8 cm]{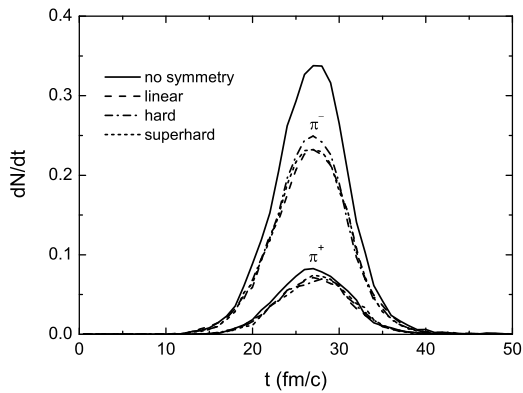}
\caption{\label{fig:epsart} Time evolution of the pion production for the cases of the linear, hard and superhard symmetry energy  and without the symmetry energy in the reaction $^{132}$Sn+$^{124}$Sn at incident energy E$_{lab}$=0.4A GeV.}
\end{figure}

\subsection{C. Influence of the in-medium pion dispersion relation}

Propagation of pion in the compressed hadronic matter is a
complicated process that is involved in the interaction of pions and
nucleons or pions and resonances. Furthermore, the pion in-medium
effect has a non-negligible influence on the charged pion ratios.
Calculations based on the QMD model show that the transverse
momentum spectrum is not influenced with an inclusion of the
$\Delta$-hole potential \cite{Fu97}. Opposite in the BUU
calculations with the $\Delta$-hole potential, the production of
pion is enhanced at the low kinetic energies \cite{Xi93}. In Fig. 7
we show the influence of the Coulomb potential and the symmetry
energy in the mean field as well as the pion optical potential in
the nuclear medium on the distributions of the transverse momenta of
the charged and neutral pion production in the left and right
panels, respectively, and also compared with the experimental data by FOPI collaboration \cite{Pe97}. We get the in-medium pion dispersion relation
by using the phenomenological approach in Eqs (24)-(27). The
propagation of pion in the nuclear matter is governed by the Coulomb
interaction and the pion optical potential. Here we use the linear
($\gamma_{s}$=1) symmetry energy and the potential parameters in Eqs (12)-(15) are got from the Skyrme force SLy6 in table 1.
The evolutions of produced pions in the baryonic matter are governed by the optical potential and also the Coulomb interaction of the charged pions and baryons.
One can see that the Coulomb potential and the symmetry energy as
shown in Eqs (9) and (20) have minor influence on the $p_{t}$
spectrum, especially for the $\pi^{+}$ and $\pi^{0}$ production. The Coulomb interaction
reduces the high-momentum $\pi^{-}$ production. The optical potential enhances the high-momentum pion
production and also the $\pi^{+}$ in the region of low $p_{t}$ comparing with the cases of the Coulomb potential and
the symmetry energy. The same phenomena is observed in the QMD
calculations for the $\pi^{0}$ transverse momentum spectrum \cite{Fu97}. The
behavior can be understood by the strong attractive potential which
enforces the pions to follow the trajectories of nucleons. Most
pions get bound by the stopped participant matter resulting in an enhancement of the low $p_{t}$ yield.
On the other hand, pions which are bound by the spectator matter are driven out
to high transverse momenta by the nucleonic flow and thus the high $p_{t}$ range is strongly enhanced. The different trends
of the $\pi^{-}$ and $\pi^{+}$ are due to the Coulomb interaction of
the charged pions and baryons. Calculations without the pion potential underpredict the high $p_{t}$ pion production is related to the missed resonances higher than N$^{\ast}$(1440). The high-$p_{t}$ pion yields are described without introducing pion potential by the calculations of Ref. \cite{La01} which takes into account an extended set of baryonic resonances with masses below 2 GeV together with the in-medium reduced NN$\rightarrow$NR cross sections.

\begin{figure*}
\includegraphics{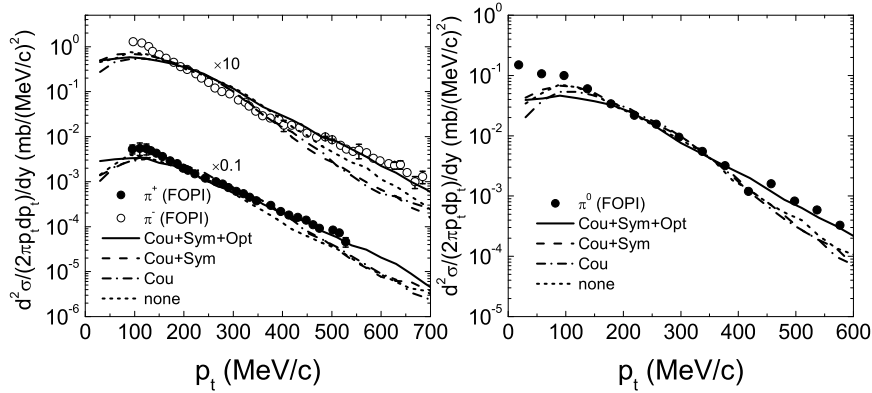}
\caption{\label{fig:wide} Comparison of the transverse momentum distributions of
charged pions and $\pi^{0}$ in the reaction $^{197}$Au+$^{197}$Au
at incident energy E$_{lab}$= 1A GeV for the cases of including all,
parts and none of the Coulomb, symmetry energy and the pion optical
potential with the available experimental data \cite{Pe97}.}
\end{figure*}

Figure 8 is a comparison of the longitudinal rapidity distribution with the same cases as shown
in Fig. 7 for the reaction $^{197}$Au+$^{197}$Au collisions at
incident energy E$_{lab}$= 1A GeV. A wider distribution is observed for the $\pi^{-}$
in the case of the optical potential. It is resulted from the fact that the optical potential enhances the high-energy
pions especially for the $\pi^{-}$ owing to the strongly attractive
potentials. The optical potential underpredicts the mid-rapidity charged pion production. The symmetry energy reduces the mid-rapidity $\pi^{-}$ production. The influence of the Coulomb potential and the symmetry
energy on the $\pi^{+}$ spectra is slightly. We also
calculated the total pion production and the $\pi^{-}$/$\pi^{+}$
yields as a function of the incident energies at the cases of the
influence of the Coulomb, symmetry energy and pion optical
potential, and also compared with the FOPI data \cite{Re07} as shown in Fig. 9. One can
see that the total pion multiplicity is not changed obviously by including any one or all parts of these
terms. However, the $\pi^{-}$/$\pi^{+}$ ratio is sensitive to the Coulomb potential and the symmetry energy owing to their sensitive dependence of the isospin degree of freedom of colliding systems, and weakly depends on the optical potential. The lower values of the $\pi^{-}$/$\pi^{+}$ ratios after including the in-medium pion dispersion at near threshold energies are consistent with the analysis in Ref. \cite{Xu10}.
Reasonable inclusion of the in-medium pion optical potential in the transport models is still an open problem, which is of importance not only in analyzing
experimental data of the pion spectrum, also in constraining the high-density behavior of the nuclear symmetry energy.

\begin{figure*}
\includegraphics{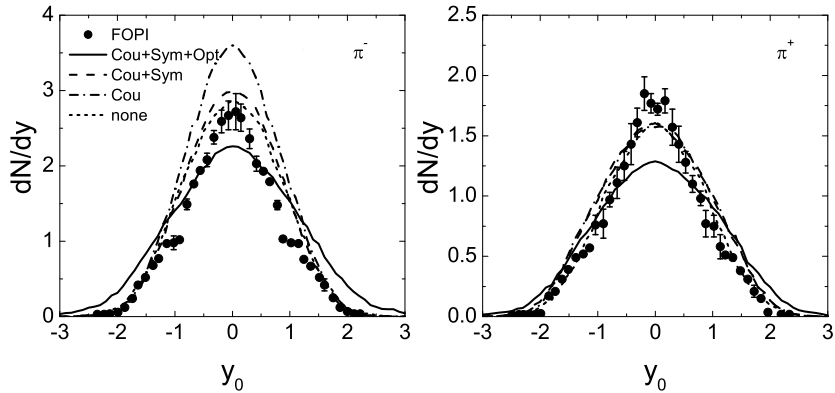}
\caption{\label{fig:wide} The same as in Fig. 7, but for the rapidity
distributions.}
\end{figure*}

\begin{figure*}
\includegraphics{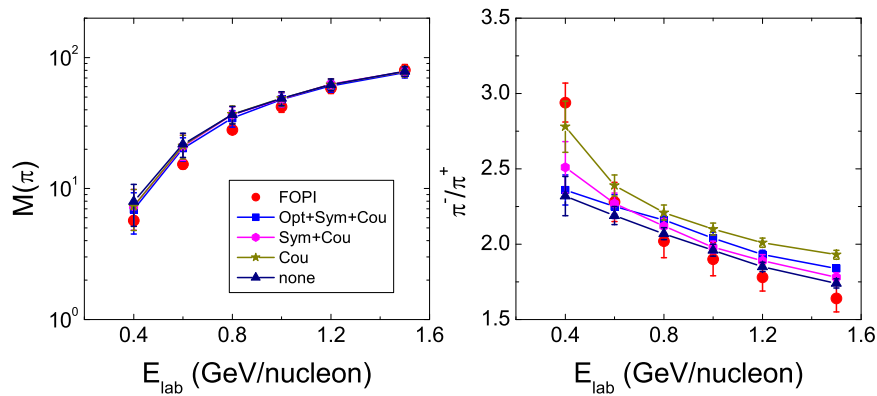}
\caption{\label{fig:wide} (Color online) The excitation functions of
the calculated pion multiplicity and the $\pi^{-}$/$\pi^{+}$ ratios
in the reaction $^{197}$Au+$^{197}$Au for head on collisions with
including all, parts and none of the Coulomb, symmetry energy and
in-medium pion optical potential, and compared with the FOPI data \cite{Re07}.}
\end{figure*}

\subsection{D. Flow distributions}

It is well known that the EoS at extreme conditions, the
in-medium properties and the nuclear dynamics in heavy-ion collisions have been widely
studied through the analysis of collective flow in experimentally associated with transport models, such as nucleonic flow, light
particles flow, and meson flow etc \cite{Re07,Gu84,Ra99}. It is possible to reconstruct the reaction plane
with flow analysis and hence to study azimuthal correlations. The analysis of the pion collective flow would be possible to get the information of
the dense isospin asymmetric nuclear matter formed in heavy-ion collisions. The
directed flow $v_{1}$ is related to the azimuth $\phi$ as
$v_{1}=\langle p_{x}/p_{t} \rangle = \langle \cos\phi \rangle$, and the elliptic flow
$v_{2}=\langle (p_{x}/p_{t})^{2}-(p_{y}/p_{t})^{2} \rangle = \langle \cos(2\phi) \rangle$, where
the angle brackets indicate averaging over all simulation events and the transverse momentum $p_{t}=\sqrt{p_{x}^{2}+p_{y}^{2}}$.
We compared the directed flows of the charged pions with protons and also with the FOPI data \cite{Re07} as a function of the longitudinal rapidity in
the $^{197}$Au+$^{197}$Au reaction at incident energy E$_{lab}$=1.5A GeV as shown in Fig. 10. The rapidity is scaled by the incident projectile in the center of mass (c.m.) frame as the relation $y_{0}=y/y_{p}$. The experimental data are selected in terms of the scaled impact parameter $b_{0}=b/b_{\textrm{max}}$ from the measured differential cross sections using a geometrical sharp-cut approximation with $b_{\textrm{max}}=1.15(A_{P}^{1/3}+A_{T}^{1/3})$. We should note that the experimental data are sharply cut in the transverse momentum range 1.0$<u_{t0}<$4.2 scaled by projectile four-velocity in the c.m. frame $u_{p}$ as the relation $u_{t0}=u_{t}/u_{p}$ with the transverse four-velocity $u_{t}$. Calculations are performed in the near central collisions with impact parameters b=2-4 fm and in the peripheral collisions with b=8-10 fm with the linear symmetry energy and without
including the pion optical potential, but the Coulomb interaction of the charged pions and baryons is considered. The well-known 'S-shape' is clear
in the distributions of the directed flows. The diagrams rotate clockwise with increasing the impact parameter. Calculated distribution trends are consistent with the experimental results in both cases. Antiflow of $\pi^{+}$ comparing the proton at larger impact parameter appears in both of the experiments and the calculations. The phenomena is caused by the shadowing effect of participant nucleons in heavy-ion collisions and was also investigated by BUU model \cite{Li94} and IQMD \cite{Ba95}.
The difference of the flow of the charged pions is due to the Coulomb interactions. In Fig. 11 we present the transverse momentum distributions of the
directed flow of charged pions and compared the recent FOPI data \cite{Re07} in the rapidity bin $-1.8<y_{0}<0$. The ImIQMD model predicts the directed flow disappears with increasing the transverse momentum ($u_{t0}>$4). Calculated distribution trends are consistent with the experimental results in both cases, although the theoretical and experimental impact parameter ranges do not precisely match.

\begin{figure*}
\includegraphics{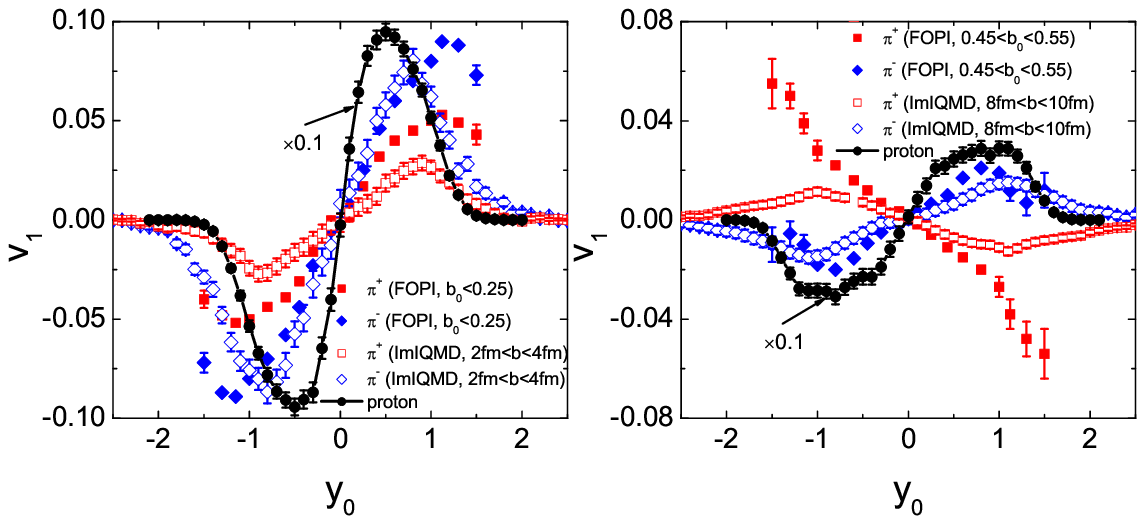}
\caption{\label{fig:wide} (Color online) Comparison the
directed flow of charged pions and protons with the FOPI data in the $^{197}$Au+$^{197}$Au reaction at incident energy E$_{lab}$=1.5A
GeV for the cases of near central and peripheral collisions.}
\end{figure*}

\begin{figure*}
\includegraphics{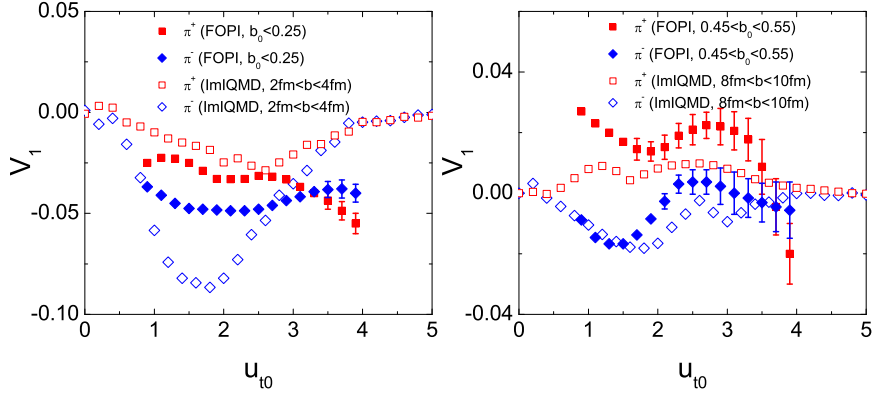}
\caption{\label{fig:wide} (Color online) Transverse momentum dependence of the
directed flow of charged pions for the system $^{197}$Au+$^{197}$Au at incident energy E$_{lab}$=1.5A
GeV for the cases of near central and peripheral collisions, and compared with the FOPI data within the rapidity bin $-1.8<y_{0}<0$.}
\end{figure*}

The elliptic flow in the $^{197}$Au+$^{197}$Au reaction is also investigated with the same framework of the directed flow as shown in Fig. 12. The flow spectra can be well reproduced in the near central collisions. However, the calculations overpredict the experimental data in the peripheral collisions owing to the different definition of the impact parameters. The structures of the flow distributions are similar. To explore more information of the isospin effect, we calculated the flow
difference of the charged pions in Fig. 13, for that the isospin difference is defined by $Dv_{1}=v_{1}^{\pi^{+}}-v_{1}^{\pi^{-}}$ for the directed flow and $Dv_{2}=v_{2}^{\pi^{+}}-v_{2}^{\pi^{-}}$ for the elliptic flow. One can see that the Coulomb potential plays a significant role in the distribution of the flow difference. The influence of the symmetry energy on the distribution is slightly for the directed flow, and relatively apparent for the elliptic flow. But the symmetry energy at any stiffness does not change the distribution structure. The dependence of flow distributions of the charged pions on the Coulomb potential and on the stiffness of the symmetry energy was also investigated by the UrQMD model and the same conclusions were reported \cite{Li06}.

\begin{figure*}
\includegraphics{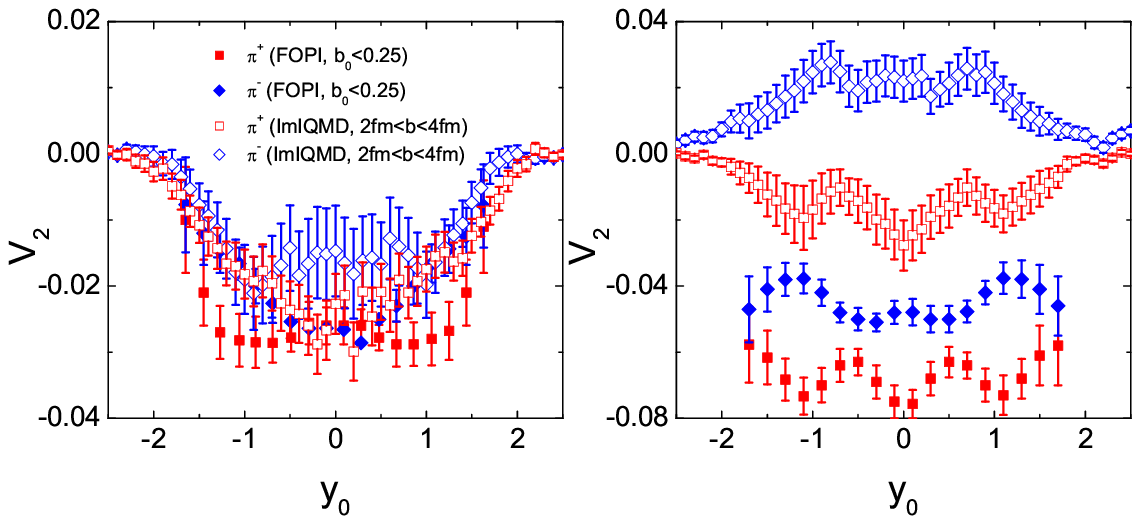}
\caption{\label{fig:wide} (Color online) The same as in Fig. 10, but for the elliptic flow.}
\end{figure*}

\begin{figure*}
\includegraphics{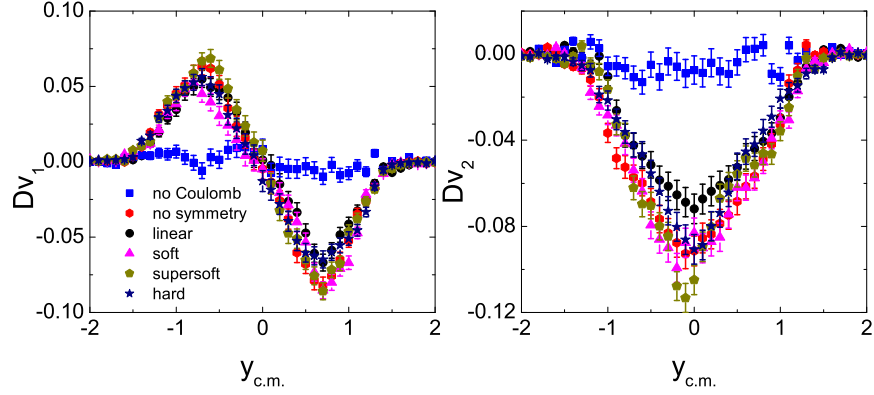}
\caption{\label{fig:wide} (Color online) Rapidity dependence of the
flow difference of the $\pi^{+}$ and $\pi^{-}$ for the system
$^{197}$Au+$^{197}$Au at incident energy E$_{lab}$=1.5A GeV and
impact parameters b=8-10 fm.}
\end{figure*}

\subsection{E. Polar anisotropy}

\begin{figure*}
\includegraphics{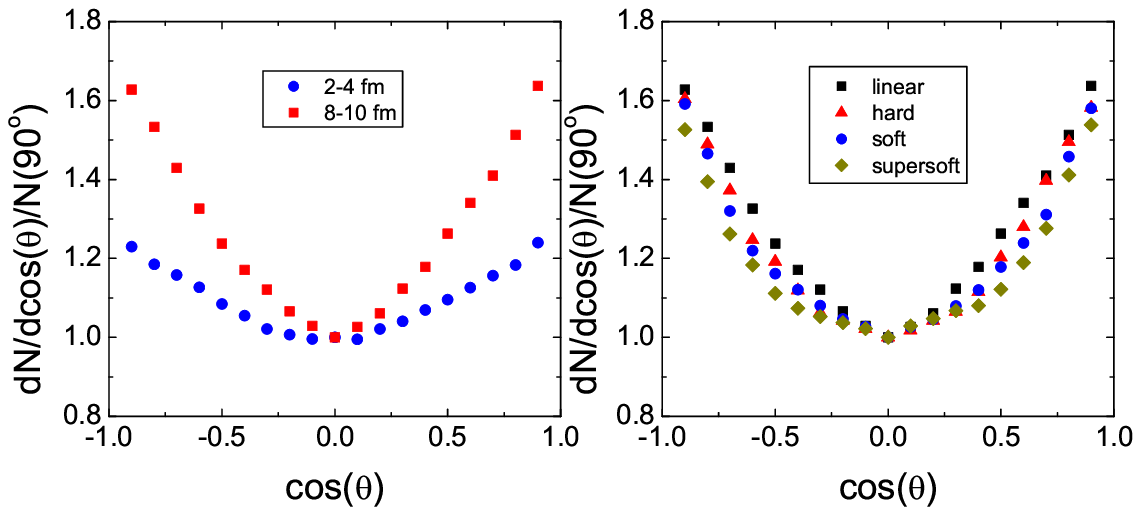}
\caption{\label{fig:wide} (Color online) Polar angle distributions of $\pi^{-}$ mesons in the reaction $^{197}$Au+$^{197}$Au at incident energy E$_{lab}$=1.5A GeV for different collision centralities (left panel) and stiffness of the symmetry energy (right panel).}
\end{figure*}

The analysis of the polar or azimuthal angle distributions of pion emission in heavy-ion collisions is useful to extract the information of the hot and dense nuclear matter. Anisotropic distribution of the polar angle of pion in center of mass system was reported by a number of experiments \cite{Re07,Ho96}. Isotropic pion emission is often assumed in the thermal model analysis of the compressed hadronic matter \cite{Av03}. We calculated the polar angle distributions of the $\pi^{-}$ in the reaction $^{197}$Au+$^{197}$Au at different collision centralities and also at the cases of the different stiffness of the symmetry energy as shown in Fig. 14. One can see that the anisotropy is pronounced in peripheral heavy-ion collisions, which is consistent with the experimental observable \cite{Re07}. The symmetry energies of the different stiffness have a minor influence on the polar angle distributions. Relatively, a wider distribution in the case of supersoft symmetry energy appears. The polar angle anisotropies result from the fact that pions are produced and emitted close to the surface of the reaction zone and they can escape without absorption by nuclear medium. An ellipsoid configuration in the reaction zone is formed in peripheral heavy-ion collisions and a significant fraction in the total pions are produced near the surface, so the large polar angle anisotropies are observed.

\section{IV. Conclusions}

The dynamics of the pion production in heavy-ion collisions in the
region of 1A GeV energies is investigated systematically by using the ImIQMD
model. The calculated total pions are consistent with the different Skyrme parameters SkP, SLy6, Ska, SIII which correspond to the
different modulus of incompressibility of symmetric nuclear matter, and the available experimental data are reproduced rather well.
The excitation functions of the $\pi^{-}/\pi^{+}$ ratio are sensitive to the stiffness of the symmetry energy. The difference is
pronounced at near threshold energies and for neutron-rich systems. Calculations show that the ratio does not monotonically depend on the stiffness of the symmetry energy at supra-saturation densities. The influence of the in-medium pion dispersion relation on the distributions of transverse momentum and longitudinal rapidity is pronounced. But it has a minor contribution on the excitation functions of the total pion production and the $\pi^{-}/\pi^{+}$ ratio. Calculations of the structure of the directed and elliptic flows are similar to the available experimental data. The flow distributions are sensitive to the Coulomb potential in the mean field, but weakly depend on the different cases of the symmetry energies. Polar angle anisotropies are pronounced in the peripheral collisions and the influence of the stiffness of the symmetry energy on the polar angle distribution is weak.

\section{Acknowledgements}

This work was supported by the National Natural Science Foundation
of China under Grant 10805061; the Special Foundation of the
President Fund; the West Doctoral Project of Chinese Academy of
Sciences; and the Major State Basic Research Development Program
under Grant 2007CB815000.

\end{document}